\documentclass[12pt]{article}
\usepackage{lingmacros}
\usepackage{tree-dvips}
\usepackage{amsmath}
\usepackage{amssymb}
\usepackage[utf8]{inputenc}
\usepackage{graphicx}
\usepackage[margin=10pt,font=small,labelfont=bf]{caption}
\usepackage[round]{natbib}

\newcommand{\captionfonts}{\normalsize}

\makeatletter  
\long\def\@makecaption#1#2{%
  \vskip\abovecaptionskip
  \sbox\@tempboxa{{\captionfonts #1: #2}}%
  \ifdim \wd\@tempboxa >\hsize
    {\captionfonts #1: #2\par}
  \else
    \hbox to\hsize{\hfil\box\@tempboxa\hfil}%
  \fi
  \vskip\belowcaptionskip}
\makeatother   

\usepackage[singlespacing]{setspace}

\title{Concerning the Neuronal Code}
\author{C. von der Malsburg\\
Frankfurt Institute for Advanced Studies}
\date{\today}
\begin{document}

\maketitle

\subsubsection*{Abstract} 

The central problem with understanding brain and mind is the neural code issue:
  understanding the matter of our brain as basis for the phenomena of our mind.
The richness with which our mind represents our environment,
  the parsimony of genetic data,
  the tremendous efficiency with which the brain learns from 
  scant sensory input and
  the creativity with which our mind constructs mental worlds
  all speak in favor of mind as an emergent phenomenon.
This raises the further issue of how the neural code
  supports these processes of organization.
The central point of this communication is 
  that the neural code has the form of structured net fragments
  that are formed by network self-organization,
  activate and de-activate on the functional time scale,  
  and spontaneously combine to form larger nets
  with the same basic structure.

\subsubsection*{The Mind-Body Problem}

While I am writing this, I am sitting on a sun-bathed terrace
  surrounded by a concrete piece of reality.
Of course this reality, to the extent that it is accessible to me,
  is a construct of my brain:
  whatever the brain doesn't represent cannot touch me.
In the normal course of affairs this figment of our brain
  is  what we take as the reality per se.
Only occasionally are we aware
  that the world out there and its reflection in us are 
  fundamentally different.
Apart from basic matters of principle, 
  we are on the one hand always restricted to a small part of the world,
  being limited to a moment and a place,
  limited moreover by our attention,
  which reflects only part of what is perceptible.
On the other hand our mind richly complements what is given by our senses
  with valuations and imaginations:
  the situation means something to us in the emotional sense
  and in the sense of opportunities to act,
  and our imagination can liberate us from place and time,
  transporting us mentally 
  into real though distant, or possible or fanciful situations.
On closer inspection it isn't even possible to make out
  a clear line between immediate reality and imagination.

We receive through our senses only scant and incomplete signals,
  shadows on the wall of Plato's cave.
To construct from these a reality 
  is possible only with the help of constitutional assumptions
  (Kant's {\it a priori}),
  with the help of masses of memory traces accumulated over the years
  (as emphasized by the empiricists),
  and with the help of extensive mental construction processes.

These construction processes are usually subconscious.
One should, however, think of the analogous and richly documented thought processes
  of mathematicians and scientists when constructing their mental edifices
  (like algebra or geometry, or the construction of geologic history 
   out of myriads of single observations)
  in order to appreciate the great importance of mental processes
  for the fabrication of our inner reality.

For essential thinkers of the 17th century the nature of the inner 
  and of the (imagined) outer reality
  --- Descartes' {\it res cogitans} and {\it res extensa} --- 
  was so different that Leibniz couldn't help attributing the collaboration 
  between his mind formulating a letter and his hand writing that letter 
  to divine intervention ("prestabilized harmony").
Spinoza, in contrast, saw brain and mind merely as different perspectives
  on the same thing.
Today, the dualism of Descartes and Leibniz is seen as left behind
  and the accepted view is essentially that of Spinoza,
  although the two perspectives
  --- that of scalpel and electrode on the one hand
      and introspection, psychophysics and psychology on the other ---
  are still so different in the mind of present-day scientists
  that they still are dualists for all intents and purposes.

Isn't now, after the preparatory work of the thinkers of the 17th, 18th and early 19th
  century and after the theoretical and experimental achievements of
  the late 19th, 20th and the incipient 21st century,
  isn't now the time to solve the mind-body problem,
  to describe the common ground behind the two perspectives,
  so that their interrelation becomes clear?
What shall we wait for?
That the solution is going to be forced on us
  by the simultaneous recording of the activity of all neurons in the waking brain
  and the complete reconstruction of its synaptic connections?
The wiring diagram of the hermaphrodite form of the nematode Caenorhabditis elegans
  has been known for many years, down to the naming of all of the 302 neurons
  and to the more than 8000 synapses \citep{White1986}
  but the behavior of the worm is still not understood on this basis.
As with perception, experimental facts about brain and mind,
  the analog to the sensory signal, are but shadows on the cave wall.
In addition to experimental data it needs effective {\it a priori} assumptions
  and active mental constructions
  in order to decipher the process behind those two perspectives, brain and mind.
Only this triad --- data, assumptions and mental construction --- can be successful.
Essential breakthroughs of science, as Maxwell's theory of electromagnetism
  or Boltzmann's statistic-mechanical explanation of the thermal phenomena,
  emphasize especially the significance of the third component,
  mental construction.

\subsubsection*{The Neural Code Issue}

I think this introduction makes in plain on what question we have to focus:
  how does the matter of my brain generate the phenomena in my mind?
How can a finite material basis generate perceptions, the representation of reality and
  a universe of imaginations, how does it engender consciousness and
  the quality of feelings?
This problem of the neural code has four inseparably intertwined aspects:
\begin{enumerate}
\item What is the nature of the state of the brain or mind at any given moment?
\item What is the nature of memory, whose structural fragments are so essential for the construction of the state?
\item What is the mechanism through which experience and thought form memory content?
\footnote{When trying to emulate the function of the brain in the computer,	more prosaic versions of these questions offer themselves:
What is the data format of brain state, what the data format of memory,
  and, what are the algorithms or what is the form of the dynamical processes
  by which those data objects are generated?
I am mentioning the computer here because
  I do believe that the questions I am discussing are 
  clearly drifting towards a crisis,
  towards the realization of artificial brains,
  designated by some as the Singularity.
The signs of this crisis are,
  first, an excitement and expectation gripping all of society
  under the names of digital revolution or artificial intelligence,
  second, the availability of the necessary computing power
    (if still at much too high cost, which is to be reduced drastically
      by a further technical revolution),
  third (driven by said attitude of expectation)
    the emergence of broad application fields for artificial brains,
  and finally, closely related to that,
    the availability of gigantic investment funds.
In my view the present situation may be likened to a huge body of water
  held back by a dam.
All that is needed to break this dam is giving correct answers 
  to the neural code questions.}
  \end{enumerate}
Answering those four questions must, of course, be guided 
  through observations of material (physiological or anatomical)
  and mental (introspective, psychological and psychophysical) kind.
Prominent is the striking contrast between the seamless unity
  we perceive in our inner being, in our consciousness, on the one hand
  and the articulation of our nervous system into a tremendous number of building elements 
  on the other. 
The mind acts like a force that generates unity in this sack full of fleas.
The non-trivial nature of this achievement is made clear
  by neurological malfunctions, which show how much our mind 
  depends on the physical components of the brain.
Malfunctions caused by local lesions have been, by the way,
  very helpful to appreciate mind as thematically articulated
  into modalities and sub-modalities
  and to identify these articulations with regions of the brain.
This localization of contents and themes has eventually been
  refined down to the level of individual neurons,
  each of which, it seems, is connected to an elementary theme or feature,
  a stimulus to which it responds by firing or a motor pattern that it triggers.

This observation gives part of an answer to the first of my questions,
  the nature of the neural code:
  the mind can be decomposed into atoms, into elementary symbols,
  and these correspond to neurons.
It is very important, however, to realize that
  as with all reduction of complex phenomena to simple building blocks
  (such as Life to molecules),
  this decomposition accomplishes only part of the task of deciphering the neural code,
  the much more complex part
  having to deal with the assembly of those elements into mental phenomena.
This is only possible in the context of answering the three other questions,
  especially those for the mechanisms generating state and memory. 

Important constraints come from observing the temporal behavior of the brain.
Preparation of spontaneous actions starts a little more than a second before execution \citep{Kornhuber,Libet}.
The reaction of the brain to new stimuli
  takes a large part of a second.
The transmission of a nervous pulse from one neuron to the next 
  takes already some milliseconds.
Well-prepared and standardized processing steps 
  take so little time \citep{RSVP,Thorpe}
  that they seem to be realized through pure feed-forward waves of neural activation.
The process in the brain is usually interpreted as a sequence of
  ``psychological moments'' \citep{block2014},
  each of which lasts one or two tenths of a second
  and can, when concentrated on, be reflected as conscious state.

According to these temporal relations the bracket between elementary processing steps 
  and whole-system reactions is very tight, 
  making clear that the generation of coherent conscious states
  is not  the result of long sequential reasoning.
We may, to be sure, also engage in long chains of reasoning, for instance
  in the context of complex mathematical proofs.
As we all know by introspection, however, these are
  the result of a slow and rather chaotic process
  in which we visit the stations of the reasoning chain 
  in random sequence, slowly putting them in order in the process.
Only sometimes, after long intensive work we can imagine
  the whole chain in front of our eyes as if it was simultaneous,
  as if we were viewing an image with deliberately shifting gaze.
Mozart is said to have once remarked that he was able to hold a whole symphony
  simultaneously in his mind,
  a statement probably to be interpreted in that same sense.
  
\subsubsection*{Brain States as Dynamical Attractors}

From such contemplations of the temporal structure and the style of processes
  in our brain we may conclude 
  that the representation of a perceived or imagined phenomenon,
  and indeed the entire conscious moment, 
  are in their nature equilibrium states
  that are constituted through the interlocking of mutually consistent forces.
This may be likened to the lattice of a crystal, in which each molecule is held
  in its position by the forces exerted by its neighbors,
  so that the whole crystal has solid rigidity
  in spite of the weakness of the individual forces.
This image of a complex arrangement stabilized by mutually consistent interactions
  seems to be highly appropriate to describe the character of the mental constructs
  that we encounter in language, music and mathematics
  or in the representation of our immediate environment.

Mathematics illustrates with particular intensity
  the stringency of the laws that rule
  the realm of self-consistent fabrics.
Thus it may be proved that besides the five Platonic bodies there can be no others,
  or that there is a fixed set of periodic crystal lattice types.
Natural numbers have a simple definition
  but their peculiar properties are to this day the subject of research.
It can be proven that the two-dimensional field of complex numbers
  cannot have a three-dimensional analog, but there is a four-dimensional one, 
  Hamilton's quaternions.
In response to this stubbornness of their subject
  mathematicians feel they are {\em discovering} pre-existing structures instead of inventing them.
These structures are entirely determined by the constraint of consistency,
  of absence of self-contradiction.
If one comes near enough to one of those fabrics, 
  it takes control and dictates the detail of its own structure.

This image of ``taking control'' is, I feel,
  highly appropriate to approach also the mechanisms 
  that are responsible for the generation of ordered brain states,
  especially of conscious moments, and of stable mental constructs in memory.
Unfortunately, mathematicians are hermetically reticent when it comes to 
  describing the history of their own discoveries
  (for a nice exception, though, see \cite{Poincare}).
One therefore has to look for physical processes putting in evidence the generation of 
  self-consistent structures.
Crystallization is, unfortunately, not appropriate for this purpose,
  as the generation of a regular molecular lattice is a very sequential process.

An instructive metaphor, however, is the B\'enard system:
  a pan, filled with some liquid like oil, heated from below.
When the heat at the bottom is slowly increased,
  so that the temperature gradient in the oil is made to surpass a certain threshold,
  the initially immobile liquid starts to agitate,
  some irregular motion patterns form,
  and under the influence of the acting forces
  (gravity, pressure, viscosity, surface tension),
  they become more and more pronounced and regular.
Under sufficiently homogeneous conditions one can see honeycomb patterns develop,
  highly regular arrays of hexagonal convection cells,
  in the center of each of which the liquid rises and near the walls of which it moves downward.
What distinguishes the winning pattern from its similar competitors
  is that it is optimal in the sense of mutual consistency of the participating forces,
  it has ``taken control'' and, amplifying itself, prevails.
 
To generate global order, forces (which typically have short range)
  have to interlink in long chains. This is bound to take time.
The temporal process of pattern formation has, however,
  nothing to do with the deterministic chain of steps of an algorithm
  or with a sequential chain of reasoning.
It rather is a chaotic sequence of immature, partially consistent patterns,
  which evolves in the sense of growing consistency
  (the analogy to an evolving eco-system is not inappropriate here at all).
  
\subsubsection*{The forces shaping memory}
  
Before we can apply the metaphor of pattern formation to the brain
  we have to return to the question regarding the nature of the neural code.
Let's first talk about the code of memory.
I follow the general conviction that memory is represented in the brain
  in the form of excitatory connections between neurons
  (assuming for simplicity that inhibitory connections 
  only play a role in housekeeping).
The strengths of excitatory connections are modified by synaptic plasticity
  (extending that concept also to the making and breaking of connections).

While for the purpose of shaping the neural activity state of the brain,
  connections are to be seen as forces
  (analogous to the viscosity etc. of B\'enard's pattern formation),
  for memory connections are {\em constituent} elements,
  that is, the material to be shaped.
What, then, are the {\em forces} between connections 
  that are responsible for forming memory?
And, what is consistency between those forces?
Before answering these questions, 
  let's briefly consider the purpose of neural connectivity patterns, of memory.
This purpose is to give shape to the rapidly changing brain states,
  making fragments of previously built imaginations available
  as components of new states.
To serve this purpose the structure of active brain states 
  has to be imprinted upon the connections
  so that memory fragments can be generated and shaped.  

Memory formation is the result of long sequences of brain states,
  which act through their statistics on the connections
  (disregarding for the moment episodic memory formation
    in hippocampus or the rather rare occurrence of imprinting).
In this process, which may be called network self-organization,
  the connectivity structure of the network on the one hand
  and the neural signal statistics on the other
  mutually adapt to better and better support each other.
Network self-organization is driven by an elementary feed-back loop:
An excitatory connection between two neurons increases
  the probability of their ``simultaneous'' (meaning, only slightly delayed) firing,
  and in return the simultaneous firing of the neurons increases the strength of the connection.
The mechanism is called Hebbian plasticity. 
This growth of synapses has to be tamed by regulatory mechanisms
  that keep the average activity of individual neurons constant
  (as described, for instance, in \cite{Triesch}).

The formative interactions between synaptic connections can, accordingly,
  be described as follows.
A single synaptic connection is too weak 
  to excite the target neuron beyond its threshold;
  for that, several fibers converging on a target neuron 
  have to fire simultaneously \citep{Abeles}. 
There will, of course, have to be a cause
  for this simultaneity of firing, 
  which ultimately is a common upstream signal source.
A decisive role in this context therefore is played by a number of signal pathways like these two 
  \[
 \begin{matrix}
  a & \rightarrow & b' \\
  \downarrow &                 & \downarrow \\
  a' & \rightarrow & b
 \end{matrix}
\] 
between neurons $a$ and $b$. 
Alternative pathways mutually increase their success in firing the target neuron,
  here $b$, 
  and are in this sense cooperative. 
On the other hand the target neuron, as remarked in the last paragraph,
  limits the number and strength of its input connections, see 
  \cite{Siddoway2014},
  which leads to competition between the fibers converging upon a neuron and the
  eventual elimination of those that are rarely active.
As a consequence, relatively few connections converge on any given neuron,
  or in other words, self-organized networks are sparse.

\begin{figure}[h!]
  \begin{center}
	\includegraphics[width=0.75\linewidth]{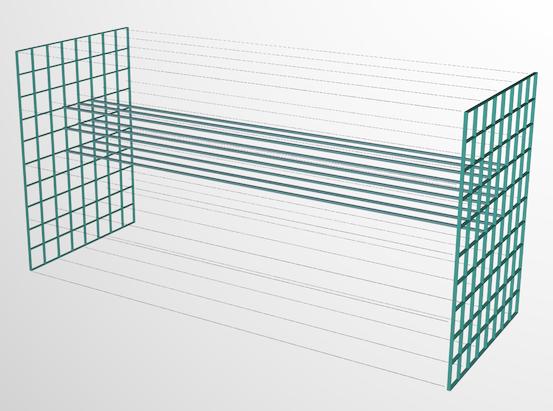}
	\renewcommand{\figurename}{Fig.}
    \caption{Two types of net structures.
    Left and right: two-dimensional neural fields.  The other type has the form of topological mappings between neural fields, mappings connecting neighboring (that is, connected) neurons in one sheet with neighboring neurons in the other (most connections pale for clarity).\hfill\hfill}
 	\label{im:nets}
  \end{center}
\end{figure}

The cerebral cortex is a richly interconnected structure in which these 
  mechanisms can play out freely.
From the above arguments it follows that cortex is dominated 
  by connectivity structures that are optimal in the sense of those two kinds of interaction
  between connections, cooperation and competition.
Such structures and their generation have been described within the visual system,  
  involving especially the ontogeny of retinotopy \citep{ProcRoySoc,rettecReview}.
These theoretical models and a number of others with the same basic structure 
  are in accordance with scores of experimental observations,
  can thus be taken as compact renderings of those experiments,
  and the underlying concepts are now common knowledge.
A prominent role is played by two-dimensionally extended nets 
  with short-range excitatory connections, so-called neural fields \citep{Cowan-Wilson,Malsburg,Amari}
  and topological (retinotopic) fiber projections between two such neural fields.
Both structures are composed of meshes of the kind described above, see Figure~\ref{im:nets},
  and are optimal in the sense of cooperation and competition. 
  
\subsubsection*{The Nature of the Neural Code}
  
Before discussing network structure any further,
  we have to revisit question Nr. 1 above,
  which we have left in a very unfinished state:
  the issue of the code of the brain state,
  of the mode of representation of live mental content.
As I wrote a long time ago \citep{Report},
  it is not enough to see the mind's state decomposed into elementary symbols,
  it also needs a means to compose those elementary symbols in hierarchically structured fashion
  so as to form more complex symbols, all the way up to the representation 
  of the currently active conscious state.
This issue has become known as the binding problem \citep{Burwick,Feldman2013}.
It will have to be solved in close coordination with the other three issues,
  the mechanism generating those complex symbols,
  the structure of memory and the mechanism of memory formation.

Here is my proposal for the form of the neural code.
The brain state is described by the set of currently active elementary symbols
  {\em and the set of currently active connections} between them.
The latter statement assumes, in radical deviation from current convictions,
  that physically existing connections can switch on and off 
  as quickly as neurons.
Connections can, however, not be activated in arbitrary combination
  but only in arrangements that have been structured by network self-organization.
Accordingly I postulate that the permanent set of cortical connections
  is an overlay of well-structured net fragments, 
  well-structured in the sense of having been created and being stabilized by self-organization. 
Active brain states are, according to this picture, generated by the activation of 
  a subset of the net fragments that make up memory.
However, net fragments cannot be activated in arbitrary combination
  but only such as to complement each other to form optimal (in the above sense) larger nets.
From now on I will imply by ``net'' or ``net fragment'' networks that are 
  near-optimal in the sense of cooperation and competition between their connections.

Let's consider the situation from the point of view of memory.
Memory consists of the set of permanent, that is, only slowly changing, connections.
Again and again, the course of time, subsets of the connections are activated.
While these are active they are plastically modified 
  in the direction of optimality, that is, towards
  improved mutual consistency.
In this sense the four neural code questions are closely intertwined:
Active brain states are part-activations of memory,
  memory is restructured by the statistics of neural activity,
  and both, brain state and memory, have the structure of nets.
  
\subsubsection*{Implementation of Rapidly Switching Connections}

I am advocating here as central element of the neural code the rapid 
  activation and de-activation of neural connections.
This is far from being a generally accepted feature of the neural paradigm
  and I would like to put out of the way possible serious reservations 
  the reader may bear against this point.
I see two mechanisms at work,
  which both have long been discussed and are based on known neurophysiological mechanisms.
  
\begin{figure}[h!]
  \begin{center}
	\includegraphics[width=0.95\linewidth]{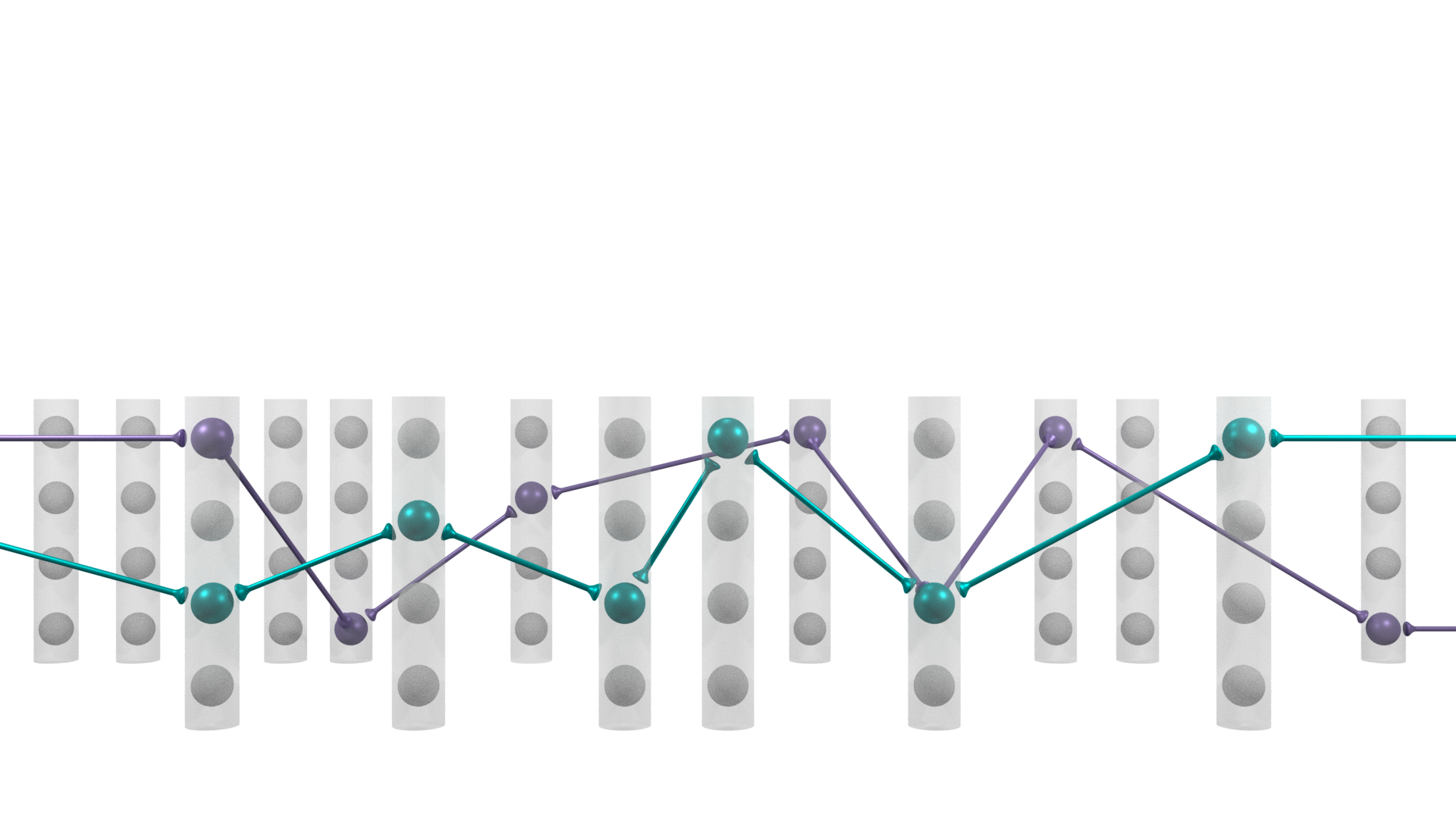}
	\renewcommand{\figurename}{Fig.}
    \caption{Overlay of Nets. 
    A two-dimensional sheet (only a one-dimensional cross section is shown for clarity)
    is populated by multicellular units (minicolumns).
    All neurons in one unit have identical external connections 
      (identical receptive field in the case of primary visual cortex).
    Inhibition within units makes sure that in a stable state only one neuron
      can be active (``winner-take-all'' inhibition).
    In a given situation, a subset of the units is actually receiving input
      (units in foreground, shown enlarged).
    As the result of learning and self-organization,
      there exists a net comprising one of the neurons in each activated unit.
    Due to this net the connected neurons win the competition within their units.
    The figure shows a second net (involving mostly units in the background).
    Different nets may share a given unit, involving different of its neurons
      (as in the third unit from the left) or even the same neuron (as in the fifth unit from the right).
    Redundant representation of features by multiple neurons per unit
      is necessary to avoid cross-talk between patterns 
      that have many features (units) in common.\hfill}
 	\label{im:mcu}
  \end{center}
\end{figure}

\begin{itemize}
\item {\it Multi-Cellular Units (see Fig.\ref{im:mcu}):} An elementary symbol, a feature, may be 
  redundantly represented by a set of alternate neurons of the same meaning.
Let's call such a set a multicellular unit.
The different neurons of the set may, however, be distinguished by their connections
  to and from neurons in other multicellular units.
At first a signal from a sensory organ (if the multi-cellular unit lies in a primary sensory area)
  or from another part of cortex will excite all neurons in the unit the same way,
  but in a second moment one of them (or a small subset) will receive ``lateral'' excitatory input
  from neurons in other units.
If an inhibitory system of the winner-take-all type 
  reigns between the neurons in a given unit,
  then the activity of all those neurons that receive little or no lateral excitation will be 
  suppressed and only one or a small subset of neurons 
  in the multicellular unit will remain active.
As result, the elementary symbols 
  that are represented by multicellular units of this structure have variable connectivity,
  realized by selective activation of neurons with corresponding permanent connections,
  see figure 2.
Such units may in the cortex be realized by the so-called minicolumns \citep{Peters}.
In the framework of associative memory this kind of structure has been proposed 
  for the purpose of de-correlating memory traces (see, for instance, \cite{Hirahara1997}).
\item {\it Control Units:} As has been proposed under the name of sigma-pi neurons \citep{sigma-pi},
  the effectiveness of a synaptic connection 
  can be controlled by other synapses that connect to the same patch of the target dendrite
  if that patch has an activation threshold \citep{Mel}.
Whole bundles of projection fibers to different neurons
  may thus be switched by individual ``control units'',
  as has been proposed by Charles Anderson \citep{VanEssenCHA}.

\end{itemize}

\begin{figure}[h!]
  \begin{center}
	\includegraphics[width=0.65\linewidth]{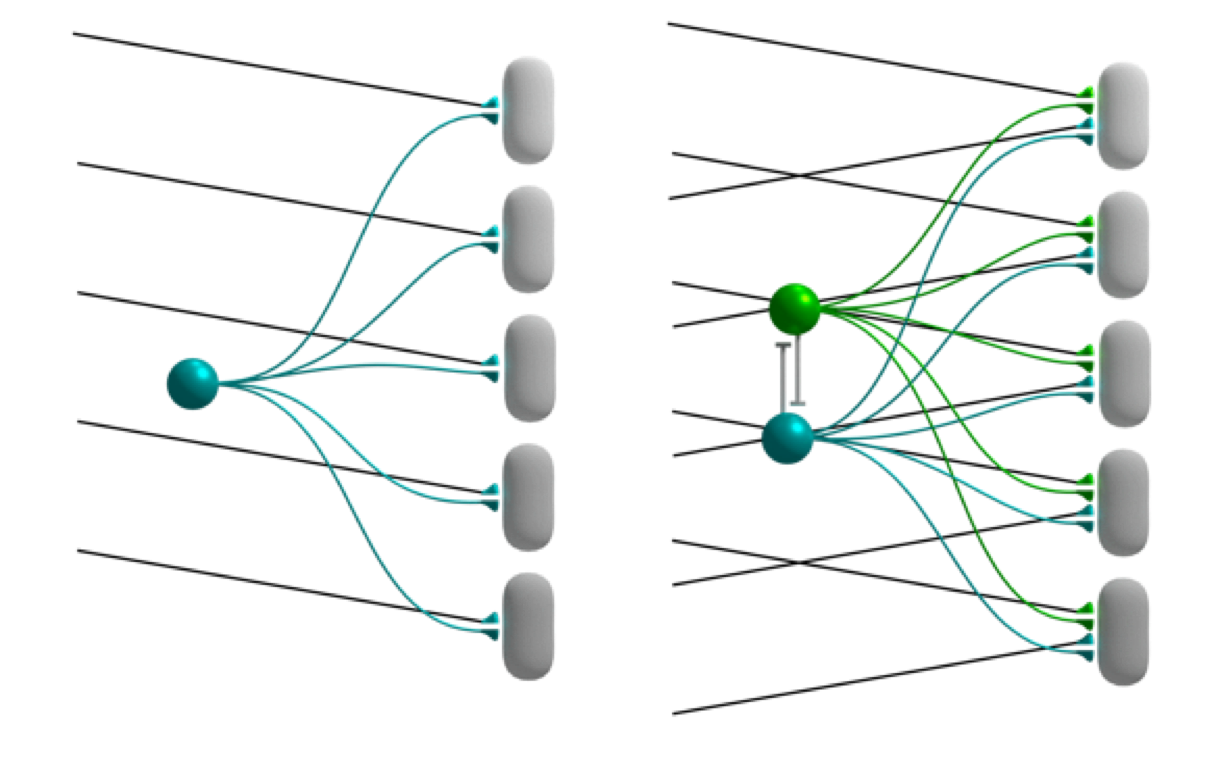}
	\renewcommand{\figurename}{Fig.}
    \caption{Control Units Gate Projection Fibers. 
    Left Panel: Synapses of projection fibers (coming from the left) 
    are co-located with synapses of control neuron (single sphere)
    on target neurons (on the right).
    Only when the control neuron is active, 
    signals of projection fibers can excite target neurons,
    due to threshold non-linearity of dendritic patches
    on which synapses of projection fibers and of the control neuron are co-located.
    Control neurons get excited in proportion to the similarity
    of the activation pattern on the projection fibers to
    the activation pattern on the target neurons
    (fibers of control neurons are assumed to conduct in both directions.)
    Right Panel: There are several (two in the figure) sets of projection fibers, 
    each of them gated by their own control unit.
    Control neurons of mutually inconsistent projection fiber sets
    (as in this case) inhibit each other,
    control neurons of mutually consistent fiber sets 
    (which form a coherent topological mapping, not shown)
    excite each other.\hfill}
 	\label{im:controlunits}
  \end{center}
\end{figure}

One may hope that within a few years the methodology of connectomics will be developed 
to the point of revealing the structure of the permanent cortical connectivity as overlay of nets. 

\subsubsection*{Significance}

How can a system that is dominated by mechanical laws,
  and for whose exposition I have only talked about form and not content,
  let alone about any relation to the environment,
  how can such a system develop intentionality and relevance to the world,
  how can it carry meaning, sentience, consciousness or express qualia?
This is, of course, the hard core of the mind-body problem:
  how can, in the sense of Spinoza, a material brain be seen as identical in essence with a mind,
  in what sense can one come to identify Descartes' {\it res extensa} with his {\it res cogitans}?

The central point to be elucidated is how the mechanics of the mind
  and the mechanics of the material building elements of the brain correspond to each other.
The movement of formal mathematics attempted something like this
  (if in rather unsatisfactory way)
  by replacing intuition by mechanical rules of symbol manipulation.
This would be more convincing if also intuition itself and the interpretation of symbols
  were realized as a formal system
  (although this would run counter to the intention of the formal mathematicians, 
  who wanted to eliminate intuition):
  but this must exactly be our aim.

It is my fundamental claim here that any kind of content and meaning
  can be represented by the proposed neural code,
  in the form of active elementary propositions
  and structured relations between them,
  that is, by graphs. 
The perhaps unfamiliar aspect is that connections
  are not restricted to the ancillary role of transmitting signals between neurons
  but are themselves dynamical variables on the time scale of brain states
  and act as active carriers of meaning in addition to their signal transmission function.

\subsubsection*{Example: Invariant Object Recognition}

Let me illustrate this with a simple example.
As we know, retinal images appear in primary visual cortex
  as two-dimensional arrays of activated neurons,
  each sensitive to a local texture element appearing in its receptive field.
The number of primary-cortical neurons is larger than the number of retinal output neurons
  by a factor of well over one hundred\footnote{The number of neurons in human primary visual cortex 
  has been estimated at 140 Million \citep{LeubaFraftsik1994} 
  and the number of human retinal output fibers at around 1 Million \citep{CurcioAllen1990}.}.
We also know that primary visual cortex abounds 
  in lateral short-range connections between its neurons.
It then seems rather natural to expect that after some learning and network self-organization
  in early life primary visual cortex assumes net structure,
  such that within the range of lateral connections
  (larger than receptive field sizes by some factor, say, five)
  all statistically significant image patch patterns
  are represented by net fragments
  that are kept separate in multi-cellular unit fashion, see Figure~\ref{im:mcu}.
Over larger distances these fragments cover images as a seamless mosaic,
  thus representing the image of an object as ``image net''.

In order to recognize an image, I posit, there needs to be, somewhere in the brain,
  a ``model net'' that is homeomorphic%
\footnote{ I am defining two nets to be homeomorphic if a one-to-one mapping can be found such
  that each neuron in one net maps one-to-one to a neuron of the same type in the other
  such that laterally connected neurons are mapped onto laterally connected neurons. 
  Both nets may be sub-nets of larger nets.  
  The term homeomorphism is just meant as an allusion to mathematical ideas. 
  A more formal definition will need to take into account the graded nature
  of the level of mutual consistency of nets.}
  to the image net in primary visual cortex.
The model net itself is embedded as fragment in a memory holding many such nets,
  and is held together not only by the short-range connections of the homeomorphism
  but also by additional longer-range connections giving the net global coherence.
  
\begin{figure}[h!]
  \begin{center}
	\includegraphics[width=0.65\linewidth]{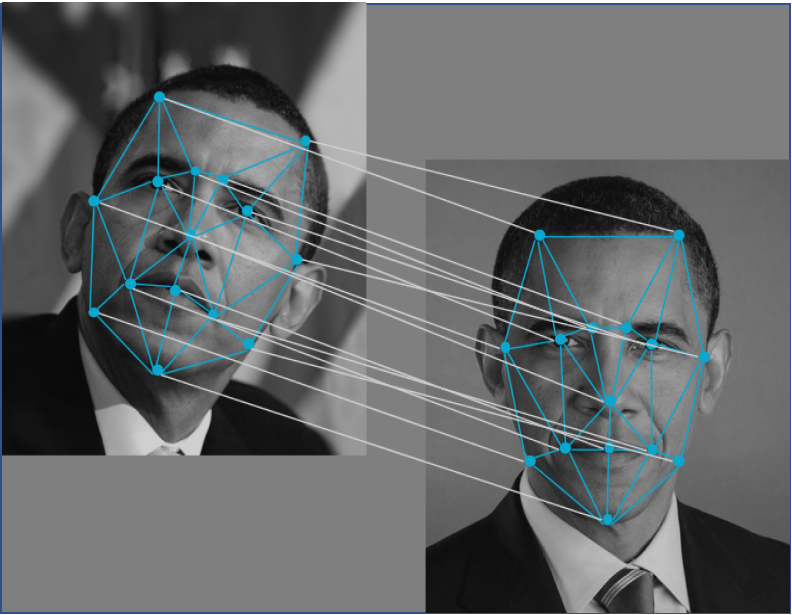}
	\renewcommand{\figurename}{Fig.}
    \caption{Example Face Recognition. 
    Left Panel: Retinal images get represented in primary visual cortex
    as two-dimensional arrays of activated neurons, each representing a local texture element
    (nodes and links symbolically stand for a much denser net). 
    Right Panel: Model of a face, in another part of the brain (fusiform complex)
    in the same basic format.
    Connecting Lines: Homeomorphic mapping, connecting feature neurons of the same type 
    and with the same connections. 
    All three networks have consistent net structure
    and they form a composite with net structure.\hfill}
 	\label{im:facerec}
  \end{center}
\end{figure}

The mappings between images and models are themselves to be realized 
  as net fragments, ``maplets'' \citep{VanEssenCHA,ZhuMal2004}.
A maplet is a set of projection fibers connecting a small patch 
  of the input sheet (primary visual cortex) 
  to a patch in the model sheet 
  (the memory containing known patterns, 
  the fusiform complex holding facial models in the human brain, for instance),
  maplet fibers connecting neighboring positions to neighboring positions.
Maplets are switched by control units,
  as proposed in \cite{VanEssenCHA}, see Figure~\ref{im:controlunits}.
Maplet control units are excited in proportion to the similarity 
    between the activity pattern on the controlled projection fibers 
    on the one hand and the activity pattern on the target neurons on the other.
Control units for competing maplets 
  (that project from different image patches to the same model patch) 
  inhibit each other, whereas control units that are consistent with each other
  (their maplets together forming a smooth topological mapping)
  excite each other.
The recognition process has to solve a chicken-and-egg problem,
  simultaneously having to home in to a definite mapping and
  activating a matching model in memory (or even composing the model from fragments).
That this is possible has been shown by simulations of 
  a complete face recognition system realized in this style see \citep{WoWoLuMa}.
For a schematic of the final composite net activated during face recognition
  see Figure~\ref{im:facerec}.

As images in retina and primary visual cortex transform with movements of objects and eye,
  a continuum of possible mappings must be held available
  to cover ranges of positions, sizes, orientations and deformations.
By composing mappings between whole figures 
  out of relatively small maplets,
  the high-dimensional space of possible mappings can be controlled
  by a finite set of control units, each one gating one of the maplets
  \citep{WolfrumMalsburg2007,Fernandes}.

As I wrote above, active connections are themselves carriers of meaning.
As illustrated in \citep{Europhysics1,Europhysics2},
  quite different images can be represented and recognized
  although they are formed by the same set of feature neurons,
  differing only in the way these neurons are connected with each other.
These connections constitute the two-dimensional extendedness of images,
  not just in the eyes of an external beholder
  but in a very effective sense: 
The net fragments of image, mapping and model can be composed to a larger net
  only on the basis of structural consistency, of homeomorphism.

Similarly, also the active projection fibers between image and model
  carry meaning, by connecting corresponding features in image and model
  and thus representing this mapping and identification.
Also the interactions between these fibers is significant.
Valid fiber projections connect neighboring neurons in one structure 
  with neighboring neurons in the other (neighborhood being defined by active connections)
  and thus consist of many meshes of the kind
 \[
 \begin{matrix}
  a & \longleftrightarrow & b \\
  \updownarrow &       & \updownarrow \\
  a' & \longleftrightarrow & b'
 \end{matrix}
\] 
(where $a$ and $a'$ are in the image and $b$ and $b'$ in the model).
These are composed of cooperating signal pathways of the kind $a\to b \to b'$ and $a\to a'\to b'$,
  while the competition between projection fibers that diverge from one point in the image
  or that converge on one point in the model is important 
  for compliance with the one-to-one structure of the mapping.
These interactions between projection fibers play the central role
  during the slow self-organization of fiber projections
  to be alternatively activated under control,
  as modeled in detail in \citep{Zhu,Bergmann,Fernandes}.
  
Such mappings make the essence of the structural relations between image and model concrete.
Already in themselves they form an indispensable kind of information,
  as argued convincingly by \citep{Arathorn} in his introduction.
The images appearing in primary visual cortex are grossly distorted
  due to the foveal inhomogeneity and other factors.
These distortions have to be compensated by the fiber projections to the model domain
  in order to re-establish the metric relations in the world.
This is possible on the basis of the {\it a priori} assumption
  that the metrics of the world are invariant to eye movements and ego motions,
  thus constituting the geometry of the visual space,
  according to Felix Klein's Erlangen program,
  as demonstrated in \citep{Zhu,Bergmann,Fernandes}.
  
By connecting the image in the primary visual cortex with an abstract model 
  (abstract in the sense of disregarding position, size and orientation of the retinal image)
  the image acquires the meaning that is implicit in the rich associations 
  of the model with the rest of the brain.
The information suppressed in the model is not lost,
  as the (bidirectional) fiber projection between model and image 
  are the basis for identifying model components with image positions,
  and the position, size and orientation of the image are encoded by the map control space,
  under alternative abstraction disregarding the structure of the image.

\subsubsection*{Schema Matching as Basis of Intelligence}

Object recognition as just described can be taken as example for a central aspect of intelligence,
  the recognition of abstract schemata in concrete situations.
Schemata have long been proposed 
  as essential ingredients of cognition and behavioral control.
Jean Piaget and~\cite{Bartlett} are often cited as important proponents of the concept.

A mind may be called intelligent if it has a large repertoire of schemata
  together with the ability to recognize and apply these in a wide array of situations.
The higher the level of abstraction of a schema the wider its scope.
The recognition of a schema in a situation
  typically suggests possible actions,
  it helps to focus attention on elements that are missing according to the schema,
  and is the basis for finding in memory other situations that are analogous in the sense of the schema. 
Schema application (though not under that name) is the central skill of a jurist,
  to be able to recognize, in a given social situation, 
  appropriate precedents or clauses in the code of law
  under which the situation can be subsumed.

A given schema may be instantiated in many concrete situations.
Conversely, a given situation may arouse several schemata, 
  and indeed only through a number of interlocking schemata can it be constituted and grasped.
Schemata may be seen as concepts
  while the representation of situations rich in detail 
  may be seen as the domain of intuition.
Concepts without intuition are empty, intuition without concepts is blind, wrote Kant.
Concepts give to perception significance and meaning,
  and conversely abstract concepts acquire content only through concrete instances.

Although the concept of the schema 
  has been discussed in the context of neural theory~\citep{Arbib,Rumelhartetal},
  its implementation in neural networks has remained elusive.
This difficulty is due to the lack of dynamic relations as part of 
  the usual data structure of neural networks.
  
Schemas have played an important role in classical artificial intelligence, 
  under various names such as frames~\citep{Minsky} or scripts~\citep{Schank}, 
  see also~\citep{Winograd},
  but their implementation has never gone beyond hand-crafted demonstrators.
Overcoming this limit needs a generic data structure for schemata, instances
  and mappings between them
  as well as a concept of self-organization to create them and let them interact.
  
Self-organized nets is the perfect medium to implement schemata and their interaction with instances.
The fundamental nature of the relation of a schema to its instance is homeomorphy. 
Let $N$ and $S$ be the nets that represent the situation and the schema, respectively.
One can speak of homeomorphy if it is possible to find a sub-net $N'$ of $N$ such
  that there is a one-to-one mapping between same-type pairs of elements in $N'$ and $S$
  fulfilling the condition that connected elements in $N'$ are mapped to connected elements in $S$.
The composite net formed by $N'$, the mapping net and $S$ will again be entirely composed
  of meshes of the kind shown above.
A homeomorphy relation is only significant if both $N$ and $S$ are sufficiently sparse networks
  (homeomorphic mappings between fully connected networks are trivial, 
  ambiguous and meaningless).
Although the general problem of sub-graph matching is intractable~\citep{NPgraph},
  mappings can be found efficiently for network structures of specific kinds
  (like planar graphs) or between labeled graphs.
Mapping between two-dimensional fields of typed neurons, 
  as in the example of object recognition~\citep{WoWoLuMa},
  or even between fields of untyped neurons~\citep{HausMals1983} 
  shows that matching can be efficient.

\subsubsection*{The Attention Focus and its Background}

Only a small sector of a visual scene is concrete and detailed
  in our inner representation.
As soon as the gaze or even attention is turned away from an object
  this object falls back onto a more abstract level,
  an object category, a position in space or a role in a functional schema.
In fact, the brain state tends to be a whole hierarchy of layers 
  from the concrete to the abstract,
  from broad context to the current focus of attention.
This goes along with a hierarchy of timescales,
  from slow (hours, even days) for the broadest context
  to fractions of a second for focal attention.
Several contexts may coexist, the mind switching between them.
It is not clear to me whether the broader contexts
  are to be counted as primed (that is, easily awakened) memory
  or as part of the active state.
It is also not clear to me whether the slower time constants
  are represented in synapses (possibly, short-lived synapses)
  or in neurons with heightened excitability or actual activity. 
These issues are relevant for the question
  of whether network self-organization and learning
  are restricted to the conscious focus of attention
  or may, to some extent, also take place in the subconscious halo.
There are many reports of learning without conscious attention,
  but more importantly we all experience that sleeping over an issue
  or putting an issue on the back burner
  may clarify things in our mind, see, e.g., \citep{Poincare}.
For a modeling study which showed improvement of functionality 
  in a ``sleep'' mode see \citep{Jitsev2010}.

It is clear, in any case, that besides the representation of context
  there is a broad halo of tentative activity.
As I will discuss below,
  sensory signals leave wide ranges of possible interpretations.
From these the mind has to sift those that are consistent
  with each other and with the context.
For this to be possible, all those tentative hypotheses
  must get excited, and out of this cloud of activity
  a small subset survives, supported by net fragments
  connecting them with each other and with the context.

\subsubsection*{Consciousness}

The conscious state is characterized by a net that comprises a critical mass of modalities.
Usually we require consciousness to describe a situation in space and time,
  including a more or less concrete agenda of the individual,
  that it defines the own situation in its relation to current events and to other persons or agents,
  and in any case that a memory protocol permits continuity in time over a hierarchy of time scales.
As I have pointed out before \citep{Kyoto},
  the essence of consciousness is the brain state's coherence in terms of content
  over a critical minimum of modalities.
What I have to add here is a definition of ``coherence in terms of content'':
  a modality-spanning consistency of signal pathways,
  meaning near-optimal net structure.

Against this background I will attempt now to approach the problem of qualia,
  the significance of the sensation of, e.g., a specific color or pain.
If indeed our mind can be described as nothing but the coming and going of nets,
  there can be only one answer to this question,
  however unsatisfied it may leave us:
When my attention is directed at a specific stimulus
  its significance is completely contained in the cascades of complex symbols, 
  represented by active nets, that are excited directly or indirectly by that stimulus.

Let's take as example the bodily sensation of a pain stimulus.
It immediately captures the attention,
  the body reacts (through reflex arcs) with evasive motions
  and possibly with systemic reactions as discharge of adrenalin,
  breaking out into a sweat, or even shock.
The pain sensation is specific in terms of quality and body location,
  it gets complemented with associations with similar sensations in the past,
  with attempts to identify a possible cause
  and with imaginations of possible harmful consequences and 
  of strategies for getting rid of the cause of the pain.
The whole event is orchestrated by a complex of behavioral schemata
  which at first are genetically determined 
  but in the course of individual development get differentiated.
The fundamental aspect of this behavioral schema is that 
  pain is the opposite of pleasure,
  that the avoidance of pain is placed high in the behavioral agenda
  and can be displaced from the position of highest degree of attention only 
  by presently more urgent matters.
All of this gives pain its substance and quality.
The essence of pain doesn't lie in the original signal of the pain fibers
  but in the cascade of activated reactions,
  such as the significance of the shot in Sarajewo lay in the World War it triggered.

Significance in brain and mind is not to be sought in specific sensory energies, 
  that is, qualities of the original signals
  as Johannes Müller saw it \citep{Mueller}.
This insight forces on us the conclusion that the specifics of sensations, qualia,
  are to be found in patterns of relations.

\subsubsection*{The Relationship between World and Brain} 

In the last section I have treated significance as something confined to the brain.
What is to be said, however, about significance in relation to the world out there,
  how can a system as described be intentional?
This question involves, again, the four questions concerning the nature and generation of the neural code.

The basis for this relation
  must be a fundamental structural trait of the world 
  that the brain has been able to capture.
The concrete mug on the table in front of me exists only once in the world.
Of my brain, however, I expect that it learns 
from the objects, structures, situations etc., encountered in the past 
to be able to interact with new situations.
That, of course, is possible only in a world that is pervaded by deep structural relations.
The possibility of describing ever new sections of a practically infinite world
  on a finite basis, a finite brain, 
  suggests that both, world and brain, are compositional (as linguists express it),
  meaning that they can be decomposed into a finite set of recurring fragments
  that combine in regular ways into hierarchies of complex structures.

\subsubsection*{Vision as Inverted Computer Graphics}

\begin{figure}[h!]
  \begin{center}
	\includegraphics[width=0.65\linewidth]{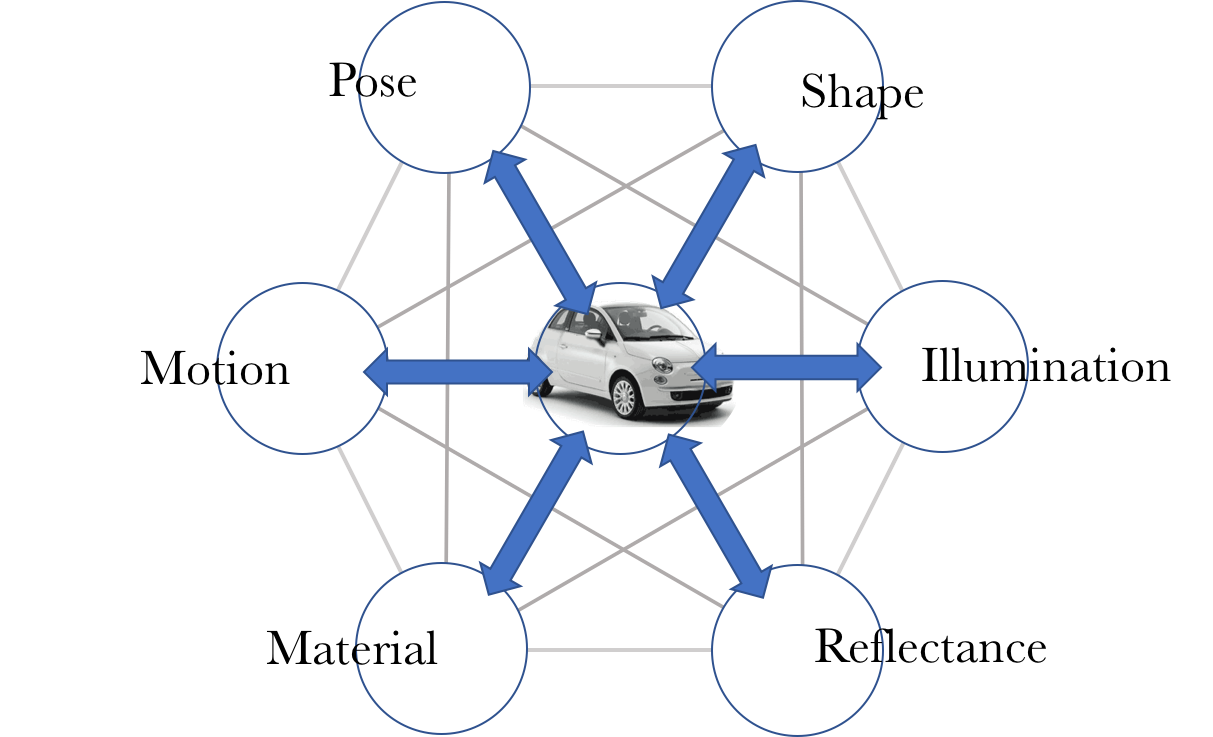}
	\renewcommand{\figurename}{Fig.}
    \caption{A Visual Architecture.
    A range of modalities (visual motion, depth, shape, reflectance, ...)
    are represented by two-dimensional sheets,
    each point of which is occupied by a local feature space
    (1D for depth, 2D for motion, 3D for color, 40D, say, for texture etc.).
    The modality sheets contain memory,
    being rich overlays of nets representing known patterns.
    The sheets are linked by ``constraint nets''
      connecting corresponding points 
      (that is, points referring to the same position on a surface in front of the eyes).
    Constraint nets represent possible feature constellations in the different modalities,
      that is, constellations that occur with statistical significance
      and have formed fragments of consistent nets.
    The visual input is highly ambiguous
    and excites many alternative feature hypotheses in the modalities.
    This cloud of possibilities collapses under two influences:
      neurons representing alternative (mutually exclusive) feature values 
      inhibit each other,
      whereas neurons standing for compatible feature values
      are connected by nets and excite each other.
    The collapse to an unambiguous interpretation of the visual input
      is driven by dynamic attraction towards a temporary global net composed 
      of many stored and mutually consistent net fragments.
    This net wins out against many less consistent net combinations
      in a process that may be called a non-serial search.\hfill}
 	\label{im:visarch}
  \end{center}
\end{figure}

Over the last two decades, computer graphics has developed the ability
  to generate very realistic-looking scenes of infinite diversity.
The structures and processes that computer graphics is using for this purpose
  may be interpreted as an ontology of the visual world.
If we were able  to invert the generative process of computer graphics,
  that is, were able to derive scene descriptions from visual input,
  we would obtain a model of visual perception, if not of cognition.

To achieve that, a fundamental difficulty has to be overcome.
While the visual scene generates sensory signals in a deterministic process
  (modeled in computer graphics),
  the inverse of this process has to search through masses of hypotheses
  to recover the right combination of scene elements.
It is as with riddles, 
  which at first confront the mind with an immense number of combinations of possible structural elements
  from which the one is to be selected that unequivocally solves the riddle. 

How does all of that translate into the language of net fragments and their interaction?
The structures and processes of computer graphics can be interpreted as net fragments
  (although here is not the place to expand on this),
  so the problem is to find an arrangement of known fragments that fit together
  to form a globally consistent net that explains the sensory input.
Explaining the input means reproducing it or,
  given unavoidable delays, predicting its continuous change.
The visual process (or more generally, perception) thus proceeds
  by first letting the sensory input create a tremendous cloud of activated neurons,
  then rapidly collapsing it, leaving active only those neurons 
  that support each other by previously learned nets.
For the schematic of a system architecture 
  that would support this process of visual perception see Figure~\ref{im:visarch}.
  
\subsubsection*{Relationship to Other Methods of Uncertainty Handling}

The uncertainty collapse just described must be seen in the context of the 
  vast literature on different methods and ideas concerning uncertainty handling.
It is too vast to be reviewed here, so let me just mention some of the relevant terms:
relaxation labeling, (deep) belief networks, Bayesian estimation, 
  Dempster-Shafer theory, particle filters, graphical models, Kalman filters,
  Hidden Markov Models or Minimal Free Energy. 
The critical issues by which these differ from each other are
  the status and the interpretation of probability, the representation of uncertainty,
  the question of how to combine evidence, issues concerning representation of state on the one hand
  and of knowledge on the other, issues concerning the acquisition of knowledge,
  and the status of a priori knowledge.
  
Common to most if not all of these conceptual frameworks is a clear distinction
  between the representation of the rapidly changing situation-dependent state
  and the representation of static or slowly changing knowledge.
The canonical version of neural networks represents the former by time-dependent neural activity
  and the latter by connections between them.
Uncertainty is represented by activity distributions over ranges of neurons
  (or in other systems over continuous idealizations of such ranges),
  and knowledge in the form of fixed connections (or generalized versions thereof) serves to 
  let mutually consistent neurons excite each other
  or inconsistent neurons inhibit each other.
  
The system I am proposing here deviates most strikingly from others
  by letting the state variables select from the knowledge pool of neural connections
  a subset to form the currently active net.
In the final, collapsed, state only those connections remain active
  that are relevant to the current situation.
The switching off of currently irrelevant connections
  frees the state from the influence of knowledge that belongs to other contexts.
The selection criterion for connections to be co-active is
  that they be consistent with each other, that is,
  minimally self-contradictory.
The efficiency of the system is due to the existence of pre-computed net fragments
  that have been shaped by slow self-organization to be self-consistent.
This requirement of self-consistency is a tremendously powerful {\it a priori} assumption
  absent from other systems.
 
Like in other systems, uncertainty is represented 
  by the superposition of neurons representing competing hypotheses,
  but here also by the superposition of connections (represented by neurons as well, see 
  the section Implementation of Rapidly Switching Connections).
This cloud of activity is collapsed,
  by inhibition and dynamic instability,
  leaving at the end a net of neurons and connections that support each other,
  meaning that they are consistent with each other in the light of previously assembled knowledge.
As to the acquisition of knowledge, see the following section on learning.

The system proposed here also differs markedly from other systems in terms of
  its data format of state, structured nets.
This lets it differ from most others by being compositional,
  representing the required hierarchies of sub-patterns and patterns
  not in terms of high-level units, 
  each sub-pattern and pattern requiring a dedicated separate unit,
  but by linking together units of lowest rank so as to form structured net fragments.
Net fragments seamlessly merge together to form higher-level, larger fragments
  in a situation-dependent way,
  thus supporting the spontaneous formation of complex high-level representations
  for phenomena encountered for the first time.
  
Finally, one word about interpreting the uncertainty distributions
  that perception has to deal with as probabilities,
  and interpreting knowledge in terms of conditional probabilities,
  as is customary in Bayesian estimation.
Of course there are phenomena, like gambling, where probabilities have objective meaning
  and it pays to calculate them {\it lege artis}.  
There is, however, no reason to take said uncertainties as probability distributions,
  no more than giving attempts at solving a riddle probability status.
Given sufficiently rich sensory input and sufficient time,
  the end result of the perceptual process is a matter of certainty:
  explaining the input is only possible on the basis of 
  a proper model of the causes behind the sensory patterns,
  and usually it is not the most likely but the only possible way to do so.
If there is lack of sufficient sensory information 
  our brain often concocts one possible interpretation
  and lets us take it as reality, without any indication of a probability value.
When the uncertainty reaches conscious level
  we attempt to deal with probabilities but are notoriously bad at computing them properly.
I therefore see no reason to burden the process of uncertainty handling 
  in a cognitive system, natural or artificial,
  with the intricacies of probability calculus, 
  especially the need to normalize probabilities to one.

\subsubsection *{Learning}

Perception is possible only if the brain already contains appropriate structural fragments.
Deducing from a flat image the geometry of a scene, for instance,  
  requires fragments that are recognizable on the basis of their two-dimensional aspect
    but contain in addition depth information.
This poses the learning problem: 
  how do these fragments, how does this information enter the mind?
  
As the bulk of our knowledge about the world has to enter our brain through perception
  we have here a serious chicken-and-egg problem:
  no perception without pre-existing fragments, no fragment formation without perception.
Two aspects are essential for understanding the solution to this impasse:
  Proper preparation of the brain and specific structure of the nursery
  in which the individual's perception and mind are fostered.

\noindent Some structures that prepare the brain for its expected environment:
\begin{itemize}\itemsep0pt
\item[i] Important sensory organs --- retina, skin and cochlea --- are spatially structured,
  reflecting the coherence of the world.
\item[ii] Different senses react to essentially different aspects of the world.
\item[iii] Efferent signals influence body motions.
\item[iv] The individual is genetically endowed with a repertoire of abstract schemata
  and corresponding reaction patterns.
\end{itemize}

\noindent As to the structure of the nursery, the following aspects may be essential:
\begin{itemize}\itemsep0pt
\item[\it a] The environment is composed of a hierarchy of recurring structural fragments.
\item[\it b] Contact with the world is articulated in a natural way 
  by the individual's position and by time.
\item[\it c] The immediate environment tends to be static on a short time scale.
\item[\it d] Events, especially motion of individual objects, 
  interrupt the temporal and spatial continuity.
\end{itemize}

Point {\it a} is evidently fundamental for the perceptibility of the world.
It will not be so easy to be fully fathomed,
but in the meantime I am basing this essay on the central thesis
  that all accessible aspects of the world 
  can be described in the brain by nets, by coherent fabrics of net fragments.

Point {\it b} is the precondition for a finite brain to grasp 
  the immensity of the world,
  by putting it through the funnel of individual situations.
As even a given situation contains too much substance,
  the brain continues the decomposition by focusing senses and attention
  and selecting and coherently representing functional components.  

The structuring of attention is supported 
  by the spontaneous occurrence of events and movements of objects (point {\it d})
  and by the movements caused by the individual (point iii),
  of the hands in front of the eyes or parallactic shifts caused by ego-motion.
Due to genetically induced mechanisms every change in the environment
  attracts attention, and all sensory signals that are simultaneous 
  with it are associated as belonging together%
\footnote{That different sensory or motor signals connected with the same external event
  are simultaneous is the original source of information on what has to do with what
  and is the archetype of temporal binding. 
Simultaneity of signals arriving at a neuron
  are also the basis for network self-organization in general.
They signal common origin in upstream events,
  though the overwhelming majority of the events driving network self-organization
  are spontaneously created inside the system.}.
Human infants and many animals stabilize the static background by staring,
  so that the immobile background can be suppressed by temporal filtering.
Moving objects escape this filtering, stand out as significant patterns
  on the basis of this ``common fate'' of their parts
  and can as such be extracted and stored from single occurrence, ({\it cf.} \cite{Loos}).
After a sufficient variety of objects have thus be recorded,
  other Gestalt rules besides common fate can be derived,
  as illustrated in basic form in \citep{Tang}.

Although sensory signals contain information about the structure of the world,
  they are also potently determined by perspective transformations.
Thus, the retinal position of the image of an object is determined by eye movements,
  its size and orientation by distance and rotation of the object or the head
  around the viewing axis
  while rotation of the object in depth alters the internal structure of the image.
Similar things are to be said about other senses, 
  for instance the influence of hand movements on tactile form perception,
  but I am restricting myself here to vision.

The topology of object surfaces is preserved by eye optics in the activity patterns of the retina,
  and due to short-range connections in the retina (point i) 
  this topology is imprinted in the form of temporal signal correlations 
  on the signals reaching the brain.
The target structures of retinal fiber projections are in natural way homeomorphic to the retinal net,
  which is the basis for ontogenetic network self-organization \citep{ProcRoySoc,rettecReview},
  resulting in retinotopic projections, 
  which in their turn maintain the topology of the received images.
One may now postulate that through a cascade of 
  dynamically switched fiber projections \citep{VanEssenCHA} 
  the images of objects are projected onto intrinsic object models,
  that is, models that represent only intrinsic properties of the objects
  and are unaffected by perspective effects.
This is Brentano's problem of intentionality,
  the problem so intensely discussed by \citep{Husserl},
  how to get from infinitely variable appearances, from the directly given sensory patterns,
  to the objects themselves.
After inspection from many angles
  it is possible to eventually construct intrinsic models of objects.
For steps in this direction see \citep{Wieghardt}. 

Here we have another chicken-and-egg problem: 
How can systematic projections be developed before there are invariant models
  and how can invariant models be generated before those projections are structured?
The problem is solved through the interplay of several basic factors.
\begin{itemize}
\item[$\alpha$] The relation between model and sensory data is structured by homeomorphy,
  by mappings that connect neurons of the same feature type under conservation of neighborhood relations.
\item[$\beta$] During the inspection of the same object under changing perspective
  (translation, rotation, scaling and rotation in depth)
  a constant model is to be put in relation to the varying input patterns
  through a set of object-independent mappings. 
\item[$\gamma$] Perspective motions (eye motions) can be initiated by the system 
  (point iii) by signals that have systematic relations to those motions 
  and the corresponding map transformations (``action-perception-cycle'',
  re-afference principle).
\item[$\delta$] The visual pathway from object to model is complemented by 
  those involving other modalities (point ii).
  Touch, in particular, helps to establish the three-dimensionality of object models.
\end{itemize}

The fundamental principle that makes it possible to solve the chicken-and-egg problem
  of simultaneously organizing fiber projections that compensate transformations and 
  invariant object models is consistency between alternative signal pathways.
In the object vision example, 
  mutual consistency has to reign 
\begin{itemize}
\item during object inspection under fixed perspective:
  between simultaneously active projection fibers 
  running between primary visual cortex and the active model,
\item during inspection of the same object under changing perspective:
  between alternative sets of projection fibers,
  which all have to support the same model with their signals, and finally
\item during inspection of different objects and scenes:
  mutual consistency has to reign within the whole system of world structure, memory structure
  and the interlocking pathways of vision and all other modalities. 
\end{itemize}  

In computer simulations we have demonstrated the simultaneous generation 
  of projection fiber systems and object models
  under exploitation of points $\alpha$ and $\beta$ 
  before birth \citep{Zhu,Bergmann} and after birth \citep{Fernandes}.

Consistency between signal pathways, 
  especially also between between changing sensory signals and model predictions,
  is the basis for our confidence in perception
  and for the naive realism that dominates our general outlook.

Let's accept, for a moment, an interpretation of active neurons
  as representing logical propositions relevant to the present situation,
  and an interpretation of the active connections 
  as arguments relevant to the present situation.
Then a conscious state would be one 
  in which all these reasoning chains are consistent with each other.
This may be compared with a self-consistent piece of mathematics
  consisting of a set of theorems connected by a web of reasoning pathways and proofs.
The degree of consistency in a brain state will never be comparable
  to the absolute stringency required in mathematics,
  but spending time observing a limited environmental scenario
  (as infants do in their stable environment)
  and thinking about it steadily increases that consistency within itself
  and with the environment.
Such imperfect but steadily increasing degree of consistency 
  is characteristic also of the grander organism
  of science, taken as a web of statements and reasoning chains,
  which over time is getting better and better in adapting itself to 
  masses of experimental observations of the studied sections of the world.    

\subsubsection*{Schema-Based Control of Behavior}

The action-perception-loop, the interplay of seeing, moving and touching,
  helps us to reconstruct the reality of our immediate environment.  
This is mainly a matter of geometry and physics.
Beyond that it is important, though, to interpret the environment as arena 
  of affordances, of opportunities for action.
This is a matter of biological significance
  and is based on a behavioral repertoire
  that has been formed during evolution,
  and some elements of which we may have inherited already 
  from our single-celled ancestors.
This repertoire is laid down by the genes in the form of abstract schemata (point iv),
  which are the essential subject of ethology.
A schema $S$, laid down as net fragment, can be identified in a scene $N$
  on the basis of homeomorphy with a part $N'$ of $N$, see above.
  and may, through further genetically determined connections, trigger a reaction.
In addition the animal may learn from the schema application
  by storing the details of the identified trigger in memory
  (as in Konrad Lorenz' instance of imprinting,
  where the gosling absorbs the image of mother-goose directly after hatching)
  in order to tune in to the actual environment.

We evidently share both the mechanisms of immediate perception and of schema-controlled behavior
  with a large number of animal species.
However, what distinguishes us is an extensive behavioral repertoire 
  directed at the organization of social interaction and communication \citep{Tomasello,Lieberman}
  and, building on that, a pyramid of schemata of immensely greater depth of abstraction,
  which we acquire through education and communication.
This enables us to expand perception to horizons way beyond the immediately accessible environment.

\subsubsection*{Conclusion}

With this essay I claim to deliver the common functional principle postulated by Spinoza
  to give rise to the very different phenomena of brain and mind.
For clarity, let me summarize my answers to the three questions
  that define the neural code issue (see the second section):
\begin{enumerate}
\item {\it What is the nature of the state of the brain or mind at any given moment?}
That state is a self-consistent net.
\item {\it What is the nature of memory?} 
Memory has the form of a connectivity pattern that is an overlay of net fragments.
\item {\it What is the mechanism through which experience and thought form memory content?}
That mechanism is network self-organization under participation of sensory input.
  \end{enumerate} 
These assumptions deviate from current ideas of the neurosciences in two respects.
First, existing connections between neural units 
  switch on and off as quickly as we think and that they are endowed with symbolic meaning
  on a par with that of neurons.
Second, the dominant factor determining the structure of active sets of connections 
  is network self-organization favoring consistency between alternative signal pathways.
The result of these modifications is a system that can be interpreted as
  describing situations by sets of currently relevant elementary assertions 
  (active neural units)
  that support each other by networks of currently relevant 
  deduction rules (active neural connections).
\bibliographystyle{plainnat}
\bibliography{NeuralCode}

\end{document}